\documentclass[10pt]{article}

\usepackage{amsxtra,amssymb,amsthm,amsmath,latexsym}

\usepackage{graphicx}

\newcommand{\R}{{\mathbb R}}

\newcommand{\bee}{\begin{equation*}}
\newcommand{\eee}{\end{equation*}}
\newcommand{\be}{\begin{equation}}
\newcommand{\ee}{\end{equation}}
\newcommand{\pn}{\par\noindent}
\title{A method for creating materials with a desired refraction 
coefficient}
\author{A G Ramm\\
\small Department of Mathematics\\[-0.8ex]
\small Kansas State University, Manhattan, KS 66506-2602, USA\\[-0.8ex]
\small \texttt{ramm@math.ksu.edu}\\
}

\begin{document} \date{} \maketitle \begin{abstract} 
It is proposed to create materials with a desired refraction coefficient
in a bounded domain $D\subset \R^3$ by embedding many small balls with 
constant refraction coefficients into
a given material. The number of small balls per unit volume around every 
point $x\in D$, i.e., their density distribution,  is calculated, as 
well as the constant refraction 
coefficients in these balls. Embedding into $D$ small balls with these
refraction coefficients according to the calculated density 
distribution creates in $D$ a material with a desired refraction 
coefficient.
\end{abstract}

\pn{\\MSC: 35J10, 74J25, 81U40; PACS 43.20.+g, 62.40.+d, 78.20.-e \\ 
{\em Key words:} scattering by small
inhomogeneities; materials; refraction coefficient; 
embedding of small inhomogeneities}

\section{Introduction}
In \cite{R509}-\cite{R536} it was proposed to create material with
a desired refraction coefficient by embedding into a given material small 
particles with  suitably chosen boundary impedances. It was proved that
any desired refraction coefficient $n^2(x)$, $\Im n^2(x)\geq 0$,
can be created in such a way in an arbitrary given bounded domain 
$D\subset \R^3$. Preparing small particle with a prescribed large 
boundary impedance may be a technologically challenging problem.
By this reason we propose in this paper a new method for
creating materials with a desired refraction coefficient.
We hope that this method may be easier to implement practically.

This  method for creating materials with a desired 
refraction coefficient  $n^2(x)$ consists of embedding into a given 
material small particles (balls) with a suitably chosen 
density distribution of the embedded particles and a suitably 
chosen constant refraction coefficients of each of the embedded 
particles.
No boundary impedances are necessary to use in
this method. Therefore, one hopes that the new method may be easier to 
implement in practice.

The density of the distribution of the embedded particles
and their constant refraction coefficients are calculated given
the desired refraction coefficient $n^2(x)$ and the refraction coefficient
 $n_0^2(x)$ of the original material in $D$.

Let us formulate the problem precisely. Assume that a bounded domain 
$D\subset \R^3$ is filled with a material with known refraction 
coefficient  $n_0^2(x)$, $\Im n_0^2(x)\geq 0$, $n_0^2(x)=1$ in
$D':=\R^3\setminus D$,  $n_0^2(x)$ is piecewise-continuous.
Throughout this paper by piecewise-continuous function we mean  
a bounded function with the set of discontinuities of
Lebesgue measure zero in $\R^3$, and do not repeat this.

The waves satisfy the equation:
\be\label{e1} L_0u_0:=[\nabla^2+k^2n_0^2(x)]u_0=0\quad in\quad \R^3,\quad
k=const>0, \ee 
\be\label{e2} u_0=e^{ik\alpha\cdot
x}+v.
\ee
Here $v$ is the scattered field, satisfying the radiation condition:
\be\label{e3}
v_r-ikv=o(r^{-1})\quad
r:=|x|\to \infty, \ee
where $\alpha\in S^2$ is the direction of the incident plane wave, 
and $S^2$ is the unit sphere in $\R^3$.
It is proved in \cite{R509} that this scattering problem under the stated
assumptions on $n_0^2(x)$ has a unique solution, and the function
$G(x,y)$, satisfying the equation 
\be\label{e4} L_0G(x,y)=-\delta(x-y)\qquad in \quad \R^3,
\ee
and the radiation condition \eqref{e3}, does 
exist and is unique.

One can write 
$$L_0=\nabla^2+k^2-q_0(x),$$ where
\be\label{eq}
q_0(x):=q_0(x,k):=k^2-k^2n_0^2(x),\qquad q_0(x)=0 \qquad x\in D'.
\ee
Let $n^2(x)$ be a desired refraction coefficient in $D$.
We assume that $n^2(x)$ is piecewise-continuous,
$\Im n^2(x)\geq 0$,  and $n^2(x)=1,\quad  x\in 
D'.$     

We wish to create material with the refraction coefficient $n^2(x)$
in $D$ by embedding into $D$ many small non-intersecting balls
$B_m$, $1\leq m \leq M$, of radius $a$, centered at the points $x_m\in 
D,$
with constant refraction coefficients $n^2_m$ in $B_m$.

Smallness of the particles means that $ka<<1$.

Let $\Delta\subset D$ be any subdomain of $D$. We assume that
the number of small particles, embedded in $\Delta$, 
is given by the 
formula: 
\be\label{e5} 
\mathcal{N}(\Delta):=\sum_{x_m\in \Delta} 1=V_a^{-1}\int_{\Delta}N(x)dx 
[1+o(1)]\qquad a\to 0,
\ee
where $N(x)\geq 0$ is a piecewise-continuous function in $D$,
and 
$$V_a:=4\pi a^3/3$$ 
is the volume of a ball of radius $a$.

Formula \eqref{e5} gives the density distribution of the centers 
of the embedded small balls in $D$. The total number of 
these balls tends to infinity as $O(V_a^{-1})=O(a^{-3})$
when  $a\to 0$.

We assume that the total volume $V(D)$ of the embedded particles
(balls) is not greater than $|D|$, where  $|D|$ is the volume of $D$, 
i.e.,
$$V(D)=V_a \mathcal{N}(D)=\int_{D}N(x)dx[1+o(1)]\leq |D|, \qquad a\to 
0.$$ 
This means physically
that although $N(x)$ can be large in some subdomains of $D$,
its average over $D$ is not greater than 1.

The scattering problem in the case of the embedded into $D$
particles is:
\be\label{e6}
(L_0+k^2\sum_{m=1}^M n_m^2 \chi_m)U=0 \qquad in \quad \R^3,
\ee
where  $\chi_m$ is the characteristic function of the ball $B_m$, i.e.,  
$\chi_m=1$ in $B_m$, $\chi_m=0$ in $B_m':=\R^3\setminus B_m$, and
\be\label{e7} 
U=u_0+V,
\ee
where $V$ satisfies the radiation condition \eqref{e3},
$u_0$ solves the scattering problem in the absence of the embedded 
particles, i.e., when $M=0$, and 
$$ n_m^2=\nu^2(x_m).$$ 
Here
$\nu^2(x)$ is some piecewise-continuous function in $D$,
$\Im \nu^2(x)\geq 0.$ 

The solution $U(x)=U_a(x)$ to problem  \eqref{e6}-\eqref{e7} depends on 
the parameter $a$, and the number $M$ of the embedded particles depends 
also on $a$, 
$$M=O(V_a^{-1})=O(a^{-3}),$$
so $M\to \infty$ at a prescribed rate as $a\to 0$.
We are interested in the limiting behavior of  $U(x)=U_a(x)$
as $a\to 0$.  Our basic result, Theorem 1, below, says that
the limit
\be\label{el} 
\lim_{a\to 0}U_a(x):=u_e(x),
\ee
does exist and satisfies the integral equation  \eqref{e15}
in Theorem 1.

From \eqref{e6}-\eqref{e7} one gets
\be\label{e8}
U(x)=u_0(x)+k^2\sum_{m=1}^M n_m^2\int_{B_m}G(x,y)U(y)dy.
\ee
This integral equation we rewrite as
\be\label{e9}
U(x)=u_0(x)+k^2\sum_{m=1}^M n_m^2\int_{B_m}G(x,y)dy U(x_m)[1+o(1)]\qquad 
a\to 0.
\ee
Here the continuity of $U$ in $B_m$, $1\leq m \leq M$, was used.
This continuity implies
$$U(y)=U(x_m)[1+o(1)]\qquad a\to 0; \qquad y\in B_m.$$
The function $U$ is twice differentiable in $\R^3$, as follows from
\eqref{e9}, so it is continuous in $D$. 

We need  three lemmas.

{\bf Lemma 1.} {\it The following relations hold}:
\be\label{e10}
\lim_{|x-y|\to 0}|x-y|G(x,y)=\frac{1}{4\pi},
\ee
\be\label{e11}
\sup_{|x-y|\geq 0}|x-y||G(x,y)|\leq c.
\ee

By $c>0$ we denote various estimation constants.

Proof of Lemma 1 is given in Section 2.

{\bf Lemma 2.} {\it The following relations hold}:
\be
\label{e12} 
\begin{split}
\int_{|y-x_m|\leq a} |x-y|^{-1}dy&=V_a|x-x_m|^{-1}, \qquad |x-x_m|\geq 
a,\\
\int_{|y-x_m|\leq a} |x-y|^{-1}dy&=2\pi\big(a^2-
\frac{|x-x_m|^2}{3}\big), \qquad |x-x_m|\leq a.
\end{split}
\ee 

Proof of Lemma 2 consists of a direct routine calculation and is 
therefore
omitted. The result of Lemma 2 is known from the potential theory.

{\bf Lemma 3.} {\it If $f$ is piecewise-continuous in $D$ and 
the points $x_m$ are distributed in $D$ by formula 
\eqref{e5}, then the following limit exists:
\be
\label{e13}
\lim_{a\to 0} V_a\sum_{m=1}^M f(x_m)=\int_Df(x)N(x)dx,
\ee
where $N(x)$ is defined in \eqref{e5}.}

This Lemma is proved in \cite{R509}.

One has:
\be\label{e14}
\int_{B_m}G(x,y)dy=V_a G(x,x_m)[1+o(1)],  \qquad a\to 0.
\ee
 From  \eqref{e9}, \eqref{e14} and \eqref{e13} our basic result follows:

{\bf Theorem 1.} {\it There exists the limit  \eqref{el} and
\be\label{e15}
u_e(x)=u_0(x)+k^2\int_D G(x,y)N(y)\nu^2(y)u_e(y)dy.
\ee
}
Physically the limiting field $u_e$ is interpreted as the effective 
(self-consistent) field in $D$.

{\bf Corollary 1.} The functions $U(x)$ and $u_e(x)$ are twice 
differentiable in $\R^3$. The function  $u_e(x)$ solves the equation:
\be\label{e16}
Lu_e(x)=0, \qquad L:=L_0+k^2N(x)\nu^2(x),
\ee
so 
\be\label{e17}
n^2(x)=n_0^2(x)+N(x)\nu^2(x).
\ee

To prove this Corollary one applies the operator $L_0$ to
equation \eqref{e15} and uses equation  \eqref{e4}.

{\bf Conclusion:} {\it To construct a material with a desired refraction 
coefficient $n^2(x)$ one embeds small balls with radius $a$, ceneterd at 
the points $x_m$, $1\leq m\leq M$, distributed by formula \eqref{e5},
and chooses $N(x)$ and $\nu^2(x)$ so that relation \eqref{e17} holds.

The choice of $N(x)$ and $\nu^2(x)$ is therefore non-unique, because the 
relation \eqref{e17} can be satisfied by infinitely many ways. For 
example, one may fix $N(x)>0$ in $D$ and then choose 
$$\nu^2(x)=\frac{n^2(x)-n_0^2(x)}{N(x)}.$$
If $n^2(x)=0$ in a subdomain $\Delta\subset D$, then one can take
$N(x)=0$ in $\Delta$.}

In Section 2 proof of Lemma 1 is  given.

\section{Proofs}

{\it Proof of Lemma 1.} We start with the equation:
\be\label{e18}
G(x,y)=g(x,y)-\int_{D} g(x,z)q_0(z)G(z,y)dz:=g-TG,
\ee
where $q_0$ is defined in  \eqref{eq}, and 
\be\label{e19}
g(x,y)=\frac{e^{ik|x-y|}}{4\pi |x-y|}.
\ee
Equation  \eqref{e18} is of Fredholm-type in the space $X$ of
functions $\psi(x,y)$ of the form $\psi(x,y)=\frac{\phi(x,y)}{|x-y|}$,
where $\phi(x,y)$ is a continuous function of its arguments, and the norm
in $X$ is defined as $||\psi||=\sup_{x,y\in \R^3}(|x-y||\psi(x,y)|)$.

We have $||g||=\frac 1 {4\pi}$. The homogeneous equation \eqref{e18}
has only the trivial solution (see \cite{R509}), so the operator
$(I+T)^{-1}$ is bounded in $X$. Therefore, $||G||\leq c||g||=\frac 
c{4\pi}$. This implies estimate \eqref{e11}.

To prove \eqref{e10}, let us multiply \eqref{e18} by $|x-y|$ and
let $|x-y|\to 0$. One has $\lim_{|x-y|\to 0}g=\frac 1 {4\pi}$.
The integral $TG$ is bounded for all $x,y\in D$, so
$\lim_{|x-y|\to 0}(|x-y|TG)=0$.
Thus, relation \eqref{e10} follows.

Lemma 1 is proved. \hfill $\Box$

\end{document}